\documentclass[aps,twocolumn,showkeys,showpacs,preprintnumbers,prl,superscriptaddress,nofootinbib,10pt]{revtex4-2}
\usepackage{graphicx,epsf,bm,amsmath,amsfonts,amssymb,epstopdf,natbib,hyperref,color,verbatim,multirow,bm,appendix}
\usepackage{tikz}
\usetikzlibrary{arrows.meta}
\usetikzlibrary{decorations.pathmorphing}
\usepackage{ulem}
\definecolor{green2}{cmyk}{0.27, 0, 1, 0.52}
\hypersetup{colorlinks=true,urlcolor=green2,citecolor=green2,linkcolor=green2,menucolor=green2,anchorcolor=green2,filecolor=green2}
\DeclareUnicodeCharacter{2212}{-}
\DeclareUnicodeCharacter{202F}{ }
\newcommand{\mpl}{m_{\rm pl}}

\begin{document}

\title{Late-Time Oscillating Quintessence in Light of DESI}

\author{Jun-Qian Jiang}
\affiliation{Korea Astronomy and Space Science Institute, Daejeon 34055, Korea}
\email{jiang@kasi.re.kr}
\author{Mustafa A. Amin}
\email{mustafa.a.amin@rice.edu}
\affiliation{Department of Physics and Astronomy, Rice University, Houston, Texas 77005, USA}
\author{Arman Shafieloo}
\email{shafieloo@kasi.re.kr}
\affiliation{Korea Astronomy and Space Science Institute, Daejeon 34055, Korea}
\affiliation{University of Science and Technology, Daejeon 34113, Korea}

\begin{abstract}
\noindent Recent DESI baryon acoustic oscillation measurements, especially when combined with Type Ia supernova and CMB data, sharpen the case for possible low-redshift dynamics in the dark energy sector. 
We study a simple and physically transparent realization of such dynamics: a quintessence field that is Hubble frozen for most of cosmic history and starts to oscillate around its minimum recently (at a redshift $z\approx 0.1$). 
This late onset of oscillations can occur in a broad class of models where the quintessence potentials have a shallow slope away from the minimum and steepen near it.
This class of models can improve the fit relative to $\Lambda$CDM, with $\Delta\chi^2\simeq -9$, while remaining competitive with common phenomenological dark energy parameterizations with the same number of parameters.
The preference is driven mainly by the background expansion history, and near the best-fit region the resonant growth of quintessence perturbations and the associated Integrated Sachs-Wolfe (ISW) contribution remain small.
More precise low-redshift distance measurements, together with late-time probes such as the ISW effect and lensing, may help distinguish this oscillating quintessence scenario from other forms of late-time dark energy dynamics.
\end{abstract}

\maketitle
\newpage

\section{Introduction}
\label{sec:introduction}
Accelerated cosmic expansion driven by a cosmological constant ($\Lambda$) has been remarkably successful in explaining a plethora of cosmological observations \cite{Planck:2018vyg}. This $\Lambda$ corresponds to a constant dark energy density $\rho_\Lambda=\Lambda/8\pi G$. However, recent baryon acoustic oscillations measurement by the DESI collaboration \cite{DESI:2024mwx,DESI:2025zgx}, in combination with Type Ia supernova distances and CMB constraints, favor models where the effective dark energy density is decreasing at late times. Kinematically, there is a preference for decreasing the acceleration close to today, albeit, the statistical strength depends on the supernova sample and on the adopted modeling assumptions (e.g. \cite{DESI:2024mwx,DESI:2025fii,DESI:2025wyn,Efstathiou:2024xcq}).

Canonical quintessence, a scalar field rolling in a potential, provides one of the simplest controlled realizations of dynamical dark energy \cite{Ratra:1987rm,Zlatev:1998tr,Copeland:2006wr,Caldwell:2005tm}.
However, conventional slow-roll or thawing models do not fully capture the DESI-preferred evolution.
Phenomenological fits often point toward phantom-divide crossing (e.g.~\cite{DESI:2025fii}), but such behavior is difficult to realize in stable, canonical theories and may partly reflect the flexibility of the chosen parameterization rather than a robust physical signal~\cite{Keeley:2025rlg,Mishra:2026tzn}.
These considerations motivate a simple canonical mechanism that remains close to $\Lambda$CDM until very late times, but then naturally reduces the dark-energy density today.

\begin{figure}[t]
\centering
\begin{tikzpicture}[scale=1.0, line cap=round, line join=round, >={Stealth[length=6pt,width=4.5pt]}]

\definecolor{potblue}{RGB}{40,90,170}
\definecolor{trajred}{RGB}{200,70,60}
\definecolor{oscgreen}{RGB}{35,145,95}
\definecolor{guidegray}{RGB}{135,135,135}
\definecolor{trajred}{RGB}{253,128,8}
\definecolor{orange}{RGB}{251,73,8}

\draw[->, semithick] (-4.25,0) -- (4.25,0) node[below] {$\phi$};
\draw[->, semithick] (0,-0.08) -- (0,2.95) node[above] {$V(\phi)$};
\draw[green2, line width=1.3pt, smooth, samples=240, domain=-4.05:4.05]
  plot (\x,{2.35*(\x*\x)/(\x*\x+0.75)});
\draw[dashed, guidegray] (-1.05,0) -- (-1.05,1.38);
\draw[dashed, guidegray] ( 1.05,0) -- ( 1.05,1.38);
\draw[dashed, guidegray] ( -3.5,0) -- ( -3.5,2.3);
\node[black, anchor=north] at (1.05,-0.06) {$M$};
\node[black, anchor=north] at (-1.05,-0.06) {-$M$};
\node[black, anchor=north] at (-3.5,-0.06) {-$10M$};
\draw[trajred, line width=1.15pt, ->]
plot[smooth, domain=-3.5:-1.1]
(\x,{2.35*(\x*\x)/(\x*\x+0.75) + 0.3});
\node[
black,
fill=lightgray!35,
draw=none,
rounded corners=4pt,
fill opacity=0.9,
text opacity=1,
inner sep=2pt,
anchor=north west,
align=center
] at (-3.3,1.9)
{\footnotesize \shortstack[l]{early-time\\slow roll}};
\draw[trajred, line width=1.05pt, ->, smooth, samples=120, domain=-0.7:0.7]
plot (\x,{2.35*(\x*\x)/(\x*\x+0.75)+0.35});
\draw[trajred, line width=1.05pt, ->, smooth, samples=120, domain=0.5:-0.5]
plot (\x,{2.6*(\x*\x)/(\x*\x+0.8)+0.15});
\node[
black,
fill=lightgray!35,
draw=none,
rounded corners=4pt,
fill opacity=0.9,
text opacity=1,
inner sep=2pt,
anchor=north west,
align=center
] at (-0.8,2.0)
{\footnotesize \shortstack[l]{late-time\\oscillations}};
\end{tikzpicture}
\caption{Schematic scalar-field potential for late-time oscillating quintessence. The field remains nearly frozen, or slow rolls, on the shallow part of the potential and begins rapid oscillations only after reaching the steeper region near the minimum. The data prefer that the field starts oscillating at late times, crossing the potential minimum first at $a_{\rm osc}\approx 0.9$. The oscillation frequency $m\sim 10^2 H_0$ and $M\sim 10^{-2}m_{\rm pl}$.}
\label{fig:potential_cartoon}
\end{figure}
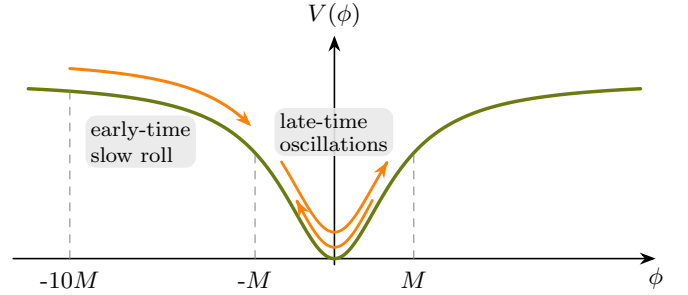

In this Letter we propose a ``simple" dark energy model to explain these recent observations: a very late-time oscillating quintessence field. The general form of the quintessence potential we have in mind is illustrated schematically in \autoref{fig:potential_cartoon}. The field remains Hubble-frozen, or slowly rolling in the flatter part of the potential for most of cosmic history. At late times, it enters a fast-rolling, oscillatory regime around the minimum. 

When confronted with recent observational data, we find the above model improves the fit significantly relative to $\Lambda$CDM, while remaining competitive with phenomenological dark energy parameterizations with the same number of additional parameters. For the best-fit region, the quintessence field starts oscillating at a redshift $z_{\rm osc}\approx 0.1$ and completes one oscillation before today. The improvement in fit has a simple origin. When the field is Hubble frozen for most of cosmic history, its energy density is nearly constant and the model remains close to $\Lambda$CDM. After the recent onset of oscillations, the dark energy density (and associated acceleration) begins to decrease naturally. 

This type of late-oscillating quintessence was studied in Ref.~\cite{Amin:2011hu}, where along with the change in the expansion rate due to the field oscillations, many rapid field oscillations were also shown to lead to resonant amplification of quintessence field fluctuations (also see \cite{Johnson:2008se}). This led to a corresponding scale-dependent growth of gravitational potentials and an enhanced Integrated Sachs-Wolfe effect \cite{Amin:2011hu}. Similar sequences of slow evolution, rapid oscillation, and possible resonance occur in other cosmological epochs, including post-inflationary dynamics and oscillating early dark energy \cite{Amin:2011hj,Lozanov:2017hjm,Poulin:2018cxd,Smith:2019ihp}. 
Related oscillatory dark energy ideas have also been explored in connection with the coincidence problem \cite{Dodelson:2000jtt}.

In the present paper, we focus on whether this canonical scalar field mechanism can produce the background dynamics preferred by observations. We also include the associated impact of quintessence perturbations, and find that they remain small in the best-fit region.

\section{Model \& Dynamics}
We consider the late-time oscillating quintessence potential introduced in Ref.~\cite{Amin:2011hu}, specialized to $\alpha=0$,
\begin{equation}
    V(\phi)=\mathcal{V}\frac{(\phi/M)^2}{1+(\phi/M)^2},\quad \mathcal{V}\equiv \frac{m^2 M^2}{2}.
\end{equation}
Here $m$ sets the curvature of the potential $V''(\phi)\approx m^2$ at its (quadratic) minimum, while $M$ controls the transition between a shallow large-field plateau and the small field quadratic minimum. Note that $\mathcal{V}$ sets the height of the potential at large field values, and $m$ sets the frequency of small amplitude field oscillations around the minimum.

While we use the above explicit form of the potential in what follows, we want to emphasize that the results are more general. In Appendix~C, we show that a broader class of scalar-field potentials with qualitative features similar to the one in Fig.~\ref{fig:potential_cartoon}, and with varying high energy physics motivations \cite{Kallosh:2013yoa,Silverstein:2008sg,Andriot:2026lac}, work equally well for our purposes. These models include cases in which the potential has non-quadratic minima ($\sim |\phi|^{2n>2}$).

For addressing the hints of late-time evolving dark energy, the dynamics we would like are ones where $\phi$ is essentially frozen during radiation and matter domination with $V(\phi_{\rm ini})\sim \mpl^2H_0^2$, and begins oscillating close to today with frequency $m\gg H_0$. This can be achieved if our quintessence field is misaligned to $\phi_{\rm ini}\gg M$ at the end of inflation and $\mathcal{V}\sim \mpl^2H_0^2$. For the field to start entering the minimum region ($\phi/M\sim 1$) close to today, we would need the initial displacement of the field $\phi_{\rm ini}/M\sim \sqrt{\mpl/M}$. Overall, we need the high energy theory to provide $m$ and $M$ such that $\mathcal{V}=m^2M^2/2\sim \mpl^2 H_0^2$, with $m\gg H_0$, while the initial misalignment must satisfy $\phi_{\rm ini}/M\sim \sqrt{m/H_0}$.

We create a collection of model-realizations by specifying $a_{\rm osc}$ and $M$, drawn from a uniform distribution over $a_{\rm osc}$ and $\log_{10}(M/\mpl)$.\footnote{In Appendix~B, we also present the results with a flat prior on the field parameters $(\log_{10}(\mathcal{V}/H^2_0 \mpl^2), \log_{10} (M/m_{\text{Pl}}))$.} Here, $a_{\rm osc}$ is the scale factor at which $\phi$ first crosses zero, approximately characterizing the onset of oscillations. For each $M$ and $a_{\rm osc}$, we adjust $\phi_{\rm ini}$ and $m/H_0$ simultaneously so that spatial flatness (correct energy density at early times) is maintained while the desired value of $a_{\rm osc}$ is reproduced. Field realizations with small $a_{\rm osc}$ undergo extremely high-frequency oscillations that make the numerical computation challenging and slow. 
Moreover, the approximation adopted below for the field perturbations may not be sufficiently accurate in this regime. To avoid these problems, we impose a lower bound $a_{\rm osc}=0.7$.

In general the quintessence field can be written as $\phi(a,\bm{x})=\phi(a)+\delta\phi(a,\bm{x})$. We modified \texttt{CLASS}~\cite{Blas:2011rf} to solve the background and (linearized) perturbation evolution.
As the homogeneous $\phi(a)$ undergoes multiple oscillations, parametric resonance leads to a scale-dependent, exponential growth in quintessence perturbations $\delta\phi(a,\bm{x})$. This growth leads to breakdown of the linearized evolution of perturbations when $\delta\phi(a,\bm{x})\sim \phi(a)$. Once this point is reached, the exponential growth is curtailed and nonlinear evolution with significant mode-coupling commences. Following Ref.~\cite{Amin:2011hu}, we define $a_{\rm nl}$,  the scalefactor when nonlinear evolution commences, as the moment when the envelope of $\max_k\sqrt{k^3P_{\delta\phi}(a,k)/(2\pi^2)}$ exceeds the envelope of the homogeneous amplitude $|\phi(a)|$. Here $ k^3P_{\delta\phi}(a,k)\sim \langle \delta\phi^2(a,\bm{x})\rangle_{L\sim k^{-1}}$ characterizes the typical variance of quintessence perturbations on the scale $k^{-1}$. 

From the Poisson equation, we can see that during the resonant growth phase of $\delta\phi$, the gravitational potential $\Psi$ also grows rapidly. For $a>a_{\rm nl}$ we freeze the gravitational potential according to $\Psi(a>a_{\rm nl},\bm{x})=\Psi(a_{\rm nl},\bm{x})$.
In practice this is implemented with two runs of \texttt{CLASS}: the first determines $a_{\rm nl}$ and the second applies the freeze-out prescription.
As we show below, the best-fit models do not in fact enter the resonant growth regime by today (i.e., $a_{\rm nl}>1$).
\section{Data}
The low-redshift data set consists of DESI DR2 BAO measurements~\cite{DESI:2025zgx}, which constrain the standard-ruler combinations 
$D_M(z)/r_d$, $D_H(z)/r_d$, and $D_V(z)/r_d$, where $D_M$ is the transverse comoving distance,
$D_H \equiv c/H(z)$ is the Hubble distance, $D_V$ is the volume-averaged distance, and $r_d$ is the sound horizon at the baryon drag epoch.
It also includes the Union3 compilation~\cite{Rubin:2023jdq} of 2087 Type Ia supernovae, which constrains the distance moduli $\mu(z)$. 
We incorporate early universe information in two ways.
First, we use the Planck PR4 \texttt{CamSpec} TT,TE,EE spectra~\cite{Rosenberg:2022sdy} with $\ell<1000$, supplemented by the PR3 low-$\ell$ TT and EE likelihoods~\cite{Planck:2019nip}. 
The high-$\ell$ spectra are excluded because they are sensitive to lensing effect, and standard prescriptions for nonlinear corrections to the matter power spectrum are not calibrated for quintessence.
As a complementary approach, we also incorporate CMB information by replacing the spectra with the compressed PR4 likelihood of Ref.~\cite{Lemos:2023xhs}, a multivariate Gaussian constraint on $(\theta_*,\Omega_bh^2,\Omega_{bc}h^2)$, where $\theta_*$ is the angular size of the sound horizon at photon decoupling, $\Omega_bh^2$ is the physical baryon density, and $\Omega_{bc}h^2\equiv(\Omega_b+\Omega_c)h^2$ is the physical density of baryons + CDM.
This prior preserves the CMB calibration of the sound horizon and physical matter densities while minimizing dependence on late-time lensing and nonlinear structure formation.

We sample the quintessence parameters along with $\{\Omega_bh^2, \Omega_{c}h^2, H_0, A_s, n_s, \tau_\text{reio}\}$\footnote{For the compressed CMB analysis, $A_s, n_s, \tau_\text{reio}$ are not sampled because the compressed likelihood only constrains the geometric and physical density parameters.} using \texttt{Cobaya}~\cite{Torrado:2020dgo} until the Gelman-Rubin convergence statistic satisfies $R-1<0.01$, and determine best-fit points with the optimizer \texttt{BOBYQA}~\cite{Cartis:2018xum,Cartis:2018jxl}.
We also used \texttt{prospect}~\cite{Holm:2023uwa} to compute the likelihood profile.

\section{Results and discussion}
\begin{figure}[t!]
  \centering
  \includegraphics[width=\linewidth]{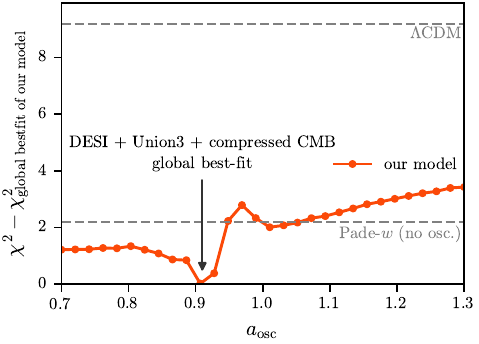}
  \caption{Likelihood profile for $a_{\rm osc}$ using DESI + Union3 + compressed CMB, shown relative to the global best-fit $\chi^2$ of our late-time oscillating quintessence model and compared with the best-fit $\Lambda$CDM and Pad\'{e}-$w$ models. The minimum at $a_{\rm osc}\simeq 0.9$ indicates that the data prefer a scenario in which the field has already entered the oscillatory regime in the recent past. The marginalized posterior distributions from the Bayesian analysis in Appendix~A, as well as the results for DESI + Union3 + CMB ($\ell<1000$), show similar behavior.}
  \label{fig:a_osc}
\end{figure}

We perform a frequentist likelihood profile analysis for DESI + Union3 + compressed CMB first, in which we fix $a_{\rm osc}$ to selected values and optimize the remaining parameters to obtain the corresponding best-fit $\chi^2$.
The result is shown in \autoref{fig:a_osc}.
It yields a best-fit with 68\% confidence interval of $a_{\rm osc}=0.916^{+0.018}_{-0.062}$.
This best-fit value of $a_{\rm osc}<1$ implies that the field entered the oscillatory regime in the past.
We also note a mild reduction in $\chi^2$ for $a_{\rm osc}$ slightly above unity, corresponding to a thawing field that is about to enter the oscillatory regime.
If the onset of oscillations is pushed further into the future, however, the quintessence dynamics approach those of $\Lambda$CDM and are correspondingly less favored.
\begin{figure}[htp]
  \centering
  \includegraphics[width=\linewidth]{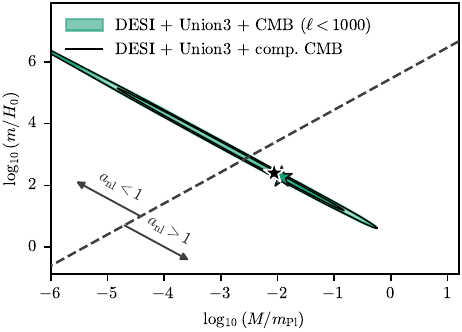}
  \caption{Marginalized posterior distributions in the $(\log_{10}(M/\mpl),\log_{10}(m/H_0))$ plane for DESI + Union3 + CMB ($\ell<1000$) and DESI + Union3 + compressed CMB. The contours show the 95\% and 68\% credible regions, and the star marks the corresponding best-fit point. The allowed region follows a narrow diagonal because reproducing the observed dark-energy scale ($\mpl^2H_0^2$) on the quintessence potential plateau ($\mathcal{V}$) approximately fixes the product $m^2M^2/2= \mathcal{V}\sim \mpl^2H_0^2$. Here $a_{\rm nl}$ indicates the scalefactor where quintessance perturbations become nonlinear.}
  \label{fig:field_parameters_2D}
\end{figure}
\autoref{fig:field_parameters_2D} shows the corresponding Bayesian marginalized posterior distribution of the field parameters in the $(\log_{10}(M/\mpl),\log_{10}(m/H_0))$ plane for the two data combinations. The allowed region follows a narrow diagonal, reflecting the fact that reproducing the observed dark energy scale on the plateau approximately fixes the product $mM$ through $\mathcal{V}=m^2M^2/2$, which should be of the same order as the critical density today.
The compressed CMB best-fit point lies near $\log_{10}(M/\mpl)\simeq -2.2$ and $\log_{10}(m/H_0)\simeq 2.5$.\footnote{The best-fit values are within an order of magnitude of those considered in a related 2011 work by one of us~\cite{Amin:2011hu}. In that case, the motivation was a ``what if'' exploration of the observational consequences of such behavior, rather than a response to DESI data. Separately, it is interesting that Ref.~\cite{Shiu:2026edl}, combining the requirement of evolving dark energy with the axionic Weak Gravity Conjecture~\cite{Heidenreich:2015wga}, obtains a similar scale, $m\sim 10^2H_0$.} The region around the best-fit point lies on the $a_{\rm nl}>1$ side of the boundary shown in the figure, where the field has not yet entered the nonlinear regime by today.
Nevertheless, regions with $a_{\rm nl}<1$ are still allowed.

\begin{figure}[t!]
  \centering
  \includegraphics[width=\linewidth]{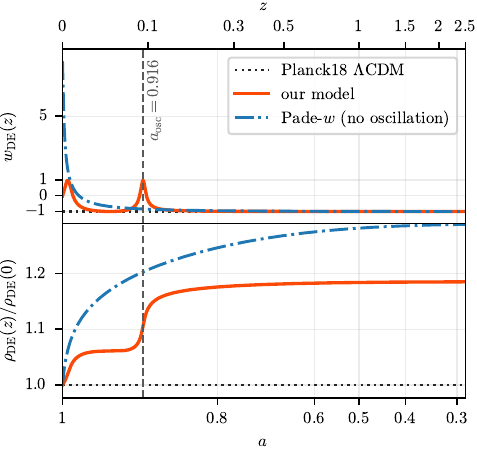}
  \caption{Top: effective equation-of-state parameter $w_{\rm DE}(z)$. Bottom: dark energy density $\rho_{\rm DE}(z)/\rho_{\rm DE}(0)$. We compare the best-fit late-time oscillating quintessence model for DESI + Union3 + compressed CMB data (red) with the best-fit Pad\'{e}-$w$ model (blue dot-dashed) and the Planck 2018 $\Lambda$CDM model (black dotted). The vertical dashed line marks the best-fit oscillation onset, $a_{\rm osc}=0.916$. }
  \label{fig:background}
\end{figure}

\begin{figure}[t!]
  \centering
  \includegraphics[width=\linewidth]{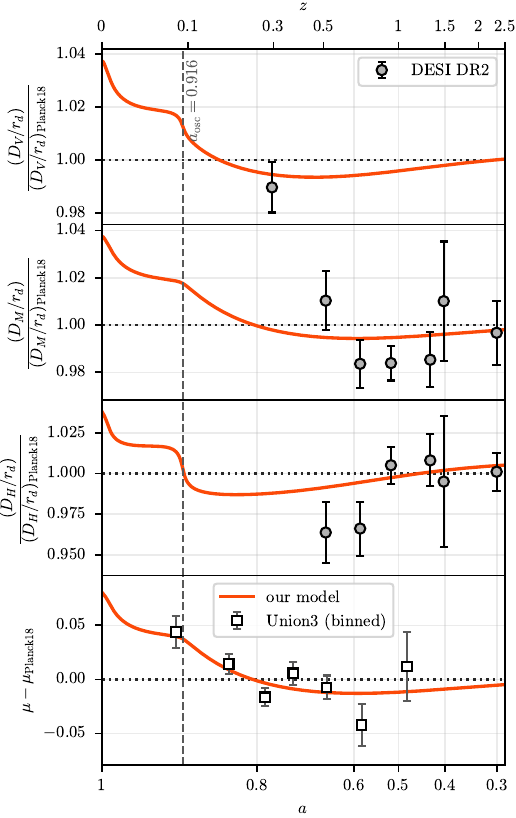}
  \caption{Corresponding distance observables for the best-fit model to DESI + Union3 + compressed CMB data, together with the DESI and Union3 data points. The Union3 sample is rebinned following the procedure of Ref.~\cite{DESI:2025zgx}. The model follows the preference of DESI and Union3 for slightly larger distances at very low redshift and slightly smaller distances at somewhat higher redshift.}
  \label{fig:observations}
\end{figure}

\autoref{fig:background} shows the late-time evolution of the background quantities at the best-fit point for DESI + Union3 + compressed CMB data.
Each time the scalar field crosses zero, the kinetic energy contribution temporarily dominates, causing the effective equation-of-state parameter $w_{\rm DE}$ to rise rapidly to $1$ before returning to $-1$.
As a result, the dark energy density $\rho_{\rm DE}$ decays in a step-like manner.
Consequently, the expansion rate $H(z)$ is relatively reduced at very low redshift for $a\gtrsim a_{\rm osc}$.
For low-redshift observations, DESI and Union3 prefer slightly larger distances at very low redshift ($z\lesssim 0.3$) than at somewhat higher redshift, see \autoref{fig:observations}.
The reduction of $H(z)$ induced by the scalar field oscillation at very low redshift naturally produces this behavior.

For DESI + Union3 + CMB ($\ell<1000$), we find that our model improves the fit relative to $\Lambda$CDM by $\Delta\chi^2=-9.32$, and $\Delta$DIC = $-4.49$\footnote{Due to the non-Gaussian and multimodal posteriors, we use the definition DIC$=\bar{\chi^2}+p_V$ with $p_V = {\rm Var}[\chi^2]/2$~\cite{Gelman2004BDA}.}.
For DESI + Union3 + compressed CMB, we find an improvement of $\Delta\chi^2=-9.19$, and $\Delta$DIC = $-3.31$.
We also test the Pad\'{e}-$w$ phenomenological dark energy parameterization on the same data sets,
\begin{equation}
    w_{\rm DE}(a)=\frac{2\epsilon_0}{2+\eta_0(a^{-3}-1)}-1,
\end{equation}
which also has two additional free parameters, $\eta_0$ and $\epsilon_0$, always satisfies $w_{\rm DE}(a)>-1$, and allows a rapid recent growth of $w_{\rm DE}$ (blue dot-dashed line in \autoref{fig:background}).
For this parameterization, the improvements relative to $\Lambda$CDM for these two datasets are $\Delta\chi^2=-7.74$ and $\Delta\chi^2=-6.64$, respectively. Our concrete quintessence model performs better than Pad\'{e}-$w$ parameterization.

For the $w_0w_a$ (Chevallier--Polarski--Linder, CPL) parameterization~\cite{Chevallier:2000qy,Linder:2002et}, which allows phantom crossing,
\begin{equation}
    w_{\rm DE}(a)=w_0+w_a(1-a),
\end{equation}
the improvements relative to $\Lambda$CDM are $\Delta\chi^2=-11.28$ and $\Delta\chi^2=-18.33$, respectively.
Our model performs similarly to the $w_0w_a$ parameterization on the DESI + Union3 + CMB ($\ell<1000$) data set, but is less favored than $w_0w_a$ for the DESI + Union3 + compressed CMB data set which has more information from high-$\ell$ part of CMB.
\footnote{When the CMB information is replaced by a BBN prior, we find comparable preferences for our model and for the $w_0w_a$ parameterization.}
This is mainly because phantom crossing allows the dark energy contribution to be reduced during the matter-dominated era.
As noted in the Introduction, however, current observations may be pointing more broadly to rapidly evolving low-redshift dark energy, rather than uniquely to phantom-divide crossing.

Before moving on to our conclusions, we reiterate that for our best-fit model, along with $m\sim 10^2H_0$ and $M\sim 10^{-2}\mpl$, we find $\phi_{\rm ini}\sim 10M\sim 0.1\mpl$. The field excursion is therefore sub-Planckian. A model building question  concerns the initial field displacement \footnote{Our quintessence field is effectively massless during inflation, so $ \phi_{\rm ini}\sim H_{\rm inf}\ll M$ if it started at $\phi=0$ before $\sim 60$ e-folds of inflation began. However, since $\mathcal{V}\sim \mpl^2H_0^2\ll H_{\rm inf}^4\ll E_{\rm inf}^4$, one could argue the field could have started anywhere.}. If the potential remains flat significantly beyond $10M$, the initial field misalignment would appear to be fine-tuned. A periodic potential with shallow regions between neighboring minima and a period of order $10M$ could ameliorate such tuning. Related discussions initial-misalignment tuning for cosine potentials can also be found in, for example, \cite{Emami:2016mrt,Shiu:2026edl}.

\section{Conclusions}
We have shown that late-time oscillating quintessence provides a simple and physically motivated realization of the low-redshift dark energy dynamics suggested by current distance data. 
In this scenario the scalar field is Hubble frozen, or slowly rolling, for most of cosmic history, so the model remains close to $\Lambda$CDM until very late times. 
Once the field reaches the steeper region near the minimum of its potential, it begins to oscillate, causing the dark energy density to decrease in a step-like manner. 
This behavior naturally reduces the expansion rate at very low redshift and thereby produces the pattern of distances preferred by the DESI BAO and Type Ia supernova data.

Using DESI DR2 BAO, Union3 supernovae, and two complementary treatments of CMB information, we find a preference for the onset of oscillations at $a_{\rm osc}\simeq 0.9$, corresponding to $z\simeq 0.1$. 
The improvement relative to $\Lambda$CDM is $\Delta\chi^2\simeq -9$. 
The same model also gives a better fit than the Pad\'e-$w$ phenomenological parameterization with the same number of additional parameters, while remaining competitive with the CPL form on the data set using CMB spectra with $\ell<1000$. 
Unlike purely phenomenological descriptions, however, the oscillating quintessence scenario arises from a canonical scalar field with a stable minimum and does not require crossing the phantom divide.
More broadly, scalar fields that end in an oscillatory phase arise in several cosmological epochs, from the end of inflation and pre-recombination early dark energy and dark matter to late-time dark energy. Despite the difference in cosmological epochs and energy scales, the similar phenomenology provides an intriguing unifying theme.

In this work, the preference for the model arises primarily from its impact on the background expansion history, especially at very low redshift; more precise future Type Ia supernova measurements can therefore provide stronger constraints.
At the perturbation level, we have treated the nonlinear evolution of the scalar field approximately. This is adequate in the region near the best fit, since the field has not yet entered the nonlinear regime there. For models in which the oscillations begin earlier, however, a more accurate nonlinear treatment in the late universe may be required (e.g.~\cite{Smith:2023fob}).
Such improvements could help to further distinguish these scenarios using observables such as the ISW effect and lensing.

\section{Acknowledgments}
\noindent 
We thank Christopher Cook and Francis-Yan Cyr-Racine for early work on this project, and Andrew Long, Mudit Jain and Scott Dodelson for helpful discussions.
M.A.\ is supported by a U.S.\ Department of Energy (DOE) Grant DE-DE-SC0010103. MA also acknowledges that part of this work was done at Aspen Center for Physics, which is supported by National Science Foundation grant PHY-2210452.
\bibliography{refs}

\begin{thebibliography}{48}%
\makeatletter
\providecommand \@ifxundefined [1]{%
 \@ifx{#1\undefined}
}%
\providecommand \@ifnum [1]{%
 \ifnum #1\expandafter \@firstoftwo
 \else \expandafter \@secondoftwo
 \fi
}%
\providecommand \@ifx [1]{%
 \ifx #1\expandafter \@firstoftwo
 \else \expandafter \@secondoftwo
 \fi
}%
\providecommand \natexlab [1]{#1}%
\providecommand \enquote  [1]{``#1''}%
\providecommand \bibnamefont  [1]{#1}%
\providecommand \bibfnamefont [1]{#1}%
\providecommand \citenamefont [1]{#1}%
\providecommand \href@noop [0]{\@secondoftwo}%
\providecommand \href [0]{\begingroup \@sanitize@url \@href}%
\providecommand \@href[1]{\@@startlink{#1}\@@href}%
\providecommand \@@href[1]{\endgroup#1\@@endlink}%
\providecommand \@sanitize@url [0]{\catcode `\\12\catcode `\$12\catcode `\&12\catcode `\#12\catcode `\^12\catcode `\_12\catcode `\%12\relax}%
\providecommand \@@startlink[1]{}%
\providecommand \@@endlink[0]{}%
\providecommand \url  [0]{\begingroup\@sanitize@url \@url }%
\providecommand \@url [1]{\endgroup\@href {#1}{\urlprefix }}%
\providecommand \urlprefix  [0]{URL }%
\providecommand \Eprint [0]{\href }%
\providecommand \doibase [0]{https://doi.org/}%
\providecommand \selectlanguage [0]{\@gobble}%
\providecommand \bibinfo  [0]{\@secondoftwo}%
\providecommand \bibfield  [0]{\@secondoftwo}%
\providecommand \translation [1]{[#1]}%
\providecommand \BibitemOpen [0]{}%
\providecommand \bibitemStop [0]{}%
\providecommand \bibitemNoStop [0]{.\EOS\space}%
\providecommand \EOS [0]{\spacefactor3000\relax}%
\providecommand \BibitemShut  [1]{\csname bibitem#1\endcsname}%
\let\auto@bib@innerbib\@empty
\bibitem [{\citenamefont {Aghanim}\ \emph {et~al.}(2020{\natexlab{a}})\citenamefont {Aghanim} \emph {et~al.}}]{Planck:2018vyg}%
  \BibitemOpen
  \bibfield  {author} {\bibinfo {author} {\bibfnamefont {N.}~\bibnamefont {Aghanim}} \emph {et~al.} (\bibinfo {collaboration} {Planck}),\ }\bibfield  {title} {\bibinfo {title} {{Planck 2018 results. VI. Cosmological parameters}},\ }\href {https://doi.org/10.1051/0004-6361/201833910} {\bibfield  {journal} {\bibinfo  {journal} {Astron. Astrophys.}\ }\textbf {\bibinfo {volume} {641}},\ \bibinfo {pages} {A6} (\bibinfo {year} {2020}{\natexlab{a}})},\ \bibinfo {note} {[Erratum: Astron.Astrophys. 652, C4 (2021)]},\ \Eprint {https://arxiv.org/abs/1807.06209} {arXiv:1807.06209 [astro-ph.CO]} \BibitemShut {NoStop}%
\bibitem [{\citenamefont {Adame}\ \emph {et~al.}(2025)\citenamefont {Adame} \emph {et~al.}}]{DESI:2024mwx}%
  \BibitemOpen
  \bibfield  {author} {\bibinfo {author} {\bibfnamefont {A.~G.}\ \bibnamefont {Adame}} \emph {et~al.} (\bibinfo {collaboration} {DESI}),\ }\bibfield  {title} {\bibinfo {title} {{DESI 2024 VI: Cosmological Constraints from the Measurements of Baryon Acoustic Oscillations}},\ }\href {https://doi.org/10.1088/1475-7516/2025/02/021} {\bibfield  {journal} {\bibinfo  {journal} {JCAP}\ }\textbf {\bibinfo {volume} {02}},\ \bibinfo {pages} {021}},\ \Eprint {https://arxiv.org/abs/2404.03002} {arXiv:2404.03002 [astro-ph.CO]} \BibitemShut {NoStop}%
\bibitem [{\citenamefont {Abdul~Karim}\ \emph {et~al.}(2025)\citenamefont {Abdul~Karim} \emph {et~al.}}]{DESI:2025zgx}%
  \BibitemOpen
  \bibfield  {author} {\bibinfo {author} {\bibfnamefont {M.}~\bibnamefont {Abdul~Karim}} \emph {et~al.} (\bibinfo {collaboration} {DESI}),\ }\bibfield  {title} {\bibinfo {title} {{DESI DR2 results. II. Measurements of baryon acoustic oscillations and cosmological constraints}},\ }\href {https://doi.org/10.1103/tr6y-kpc6} {\bibfield  {journal} {\bibinfo  {journal} {Phys. Rev. D}\ }\textbf {\bibinfo {volume} {112}},\ \bibinfo {pages} {083515} (\bibinfo {year} {2025})},\ \Eprint {https://arxiv.org/abs/2503.14738} {arXiv:2503.14738 [astro-ph.CO]} \BibitemShut {NoStop}%
\bibitem [{\citenamefont {Lodha}\ \emph {et~al.}(2025)\citenamefont {Lodha} \emph {et~al.}}]{DESI:2025fii}%
  \BibitemOpen
  \bibfield  {author} {\bibinfo {author} {\bibfnamefont {K.}~\bibnamefont {Lodha}} \emph {et~al.} (\bibinfo {collaboration} {DESI}),\ }\bibfield  {title} {\bibinfo {title} {{Extended dark energy analysis using DESI DR2 BAO measurements}},\ }\href {https://doi.org/10.1103/w4c6-1r5j} {\bibfield  {journal} {\bibinfo  {journal} {Phys. Rev. D}\ }\textbf {\bibinfo {volume} {112}},\ \bibinfo {pages} {083511} (\bibinfo {year} {2025})},\ \Eprint {https://arxiv.org/abs/2503.14743} {arXiv:2503.14743 [astro-ph.CO]} \BibitemShut {NoStop}%
\bibitem [{\citenamefont {Gu}\ \emph {et~al.}(2025)\citenamefont {Gu} \emph {et~al.}}]{DESI:2025wyn}%
  \BibitemOpen
  \bibfield  {author} {\bibinfo {author} {\bibfnamefont {G.}~\bibnamefont {Gu}} \emph {et~al.} (\bibinfo {collaboration} {DESI}),\ }\bibfield  {title} {\bibinfo {title} {{Dynamical dark energy in light of the DESI DR2 baryonic acoustic oscillations measurements}},\ }\href {https://doi.org/10.1038/s41550-025-02669-6} {\bibfield  {journal} {\bibinfo  {journal} {Nature Astron.}\ }\textbf {\bibinfo {volume} {9}},\ \bibinfo {pages} {1879} (\bibinfo {year} {2025})},\ \bibinfo {note} {[Erratum: Nature Astron. 9, 1898 (2025)]},\ \Eprint {https://arxiv.org/abs/2504.06118} {arXiv:2504.06118 [astro-ph.CO]} \BibitemShut {NoStop}%
\bibitem [{\citenamefont {Efstathiou}(2025)}]{Efstathiou:2024xcq}%
  \BibitemOpen
  \bibfield  {author} {\bibinfo {author} {\bibfnamefont {G.}~\bibnamefont {Efstathiou}},\ }\bibfield  {title} {\bibinfo {title} {{Evolving dark energy or supernovae systematics?}},\ }\href {https://doi.org/10.1093/mnras/staf301} {\bibfield  {journal} {\bibinfo  {journal} {Mon. Not. Roy. Astron. Soc.}\ }\textbf {\bibinfo {volume} {538}},\ \bibinfo {pages} {875} (\bibinfo {year} {2025})},\ \Eprint {https://arxiv.org/abs/2408.07175} {arXiv:2408.07175 [astro-ph.CO]} \BibitemShut {NoStop}%
\bibitem [{\citenamefont {Ratra}\ and\ \citenamefont {Peebles}(1988)}]{Ratra:1987rm}%
  \BibitemOpen
  \bibfield  {author} {\bibinfo {author} {\bibfnamefont {B.}~\bibnamefont {Ratra}}\ and\ \bibinfo {author} {\bibfnamefont {P.~J.~E.}\ \bibnamefont {Peebles}},\ }\bibfield  {title} {\bibinfo {title} {Cosmological consequences of a rolling homogeneous scalar field},\ }\href@noop {} {\bibfield  {journal} {\bibinfo  {journal} {Phys. Rev. D}\ }\textbf {\bibinfo {volume} {37}},\ \bibinfo {pages} {3406} (\bibinfo {year} {1988})}\BibitemShut {NoStop}%
\bibitem [{\citenamefont {Zlatev}\ \emph {et~al.}(1999)\citenamefont {Zlatev}, \citenamefont {Wang},\ and\ \citenamefont {Steinhardt}}]{Zlatev:1998tr}%
  \BibitemOpen
  \bibfield  {author} {\bibinfo {author} {\bibfnamefont {I.}~\bibnamefont {Zlatev}}, \bibinfo {author} {\bibfnamefont {L.}~\bibnamefont {Wang}},\ and\ \bibinfo {author} {\bibfnamefont {P.~J.}\ \bibnamefont {Steinhardt}},\ }\bibfield  {title} {\bibinfo {title} {{Quintessence, Cosmic Coincidence, and the Cosmological Constant}},\ }\href {https://doi.org/10.1103/PhysRevLett.82.896} {\bibfield  {journal} {\bibinfo  {journal} {Phys. Rev. Lett.}\ }\textbf {\bibinfo {volume} {82}},\ \bibinfo {pages} {896} (\bibinfo {year} {1999})},\ \Eprint {https://arxiv.org/abs/astro-ph/9807002} {arXiv:astro-ph/9807002} \BibitemShut {NoStop}%
\bibitem [{\citenamefont {Copeland}\ \emph {et~al.}(2006)\citenamefont {Copeland}, \citenamefont {Sami},\ and\ \citenamefont {Tsujikawa}}]{Copeland:2006wr}%
  \BibitemOpen
  \bibfield  {author} {\bibinfo {author} {\bibfnamefont {E.~J.}\ \bibnamefont {Copeland}}, \bibinfo {author} {\bibfnamefont {M.}~\bibnamefont {Sami}},\ and\ \bibinfo {author} {\bibfnamefont {S.}~\bibnamefont {Tsujikawa}},\ }\bibfield  {title} {\bibinfo {title} {{Dynamics of dark energy}},\ }\href {https://doi.org/10.1142/S021827180600942X} {\bibfield  {journal} {\bibinfo  {journal} {Int. J. Mod. Phys. D}\ }\textbf {\bibinfo {volume} {15}},\ \bibinfo {pages} {1753} (\bibinfo {year} {2006})},\ \Eprint {https://arxiv.org/abs/hep-th/0603057} {arXiv:hep-th/0603057} \BibitemShut {NoStop}%
\bibitem [{\citenamefont {Caldwell}\ and\ \citenamefont {Linder}(2005)}]{Caldwell:2005tm}%
  \BibitemOpen
  \bibfield  {author} {\bibinfo {author} {\bibfnamefont {R.~R.}\ \bibnamefont {Caldwell}}\ and\ \bibinfo {author} {\bibfnamefont {E.~V.}\ \bibnamefont {Linder}},\ }\bibfield  {title} {\bibinfo {title} {{The Limits of quintessence}},\ }\href {https://doi.org/10.1103/PhysRevLett.95.141301} {\bibfield  {journal} {\bibinfo  {journal} {Phys. Rev. Lett.}\ }\textbf {\bibinfo {volume} {95}},\ \bibinfo {pages} {141301} (\bibinfo {year} {2005})},\ \Eprint {https://arxiv.org/abs/astro-ph/0505494} {arXiv:astro-ph/0505494} \BibitemShut {NoStop}%
\bibitem [{\citenamefont {Keeley}\ \emph {et~al.}(2025)\citenamefont {Keeley}, \citenamefont {Shafieloo},\ and\ \citenamefont {Matthewson}}]{Keeley:2025rlg}%
  \BibitemOpen
  \bibfield  {author} {\bibinfo {author} {\bibfnamefont {R.~E.}\ \bibnamefont {Keeley}}, \bibinfo {author} {\bibfnamefont {A.}~\bibnamefont {Shafieloo}},\ and\ \bibinfo {author} {\bibfnamefont {W.~L.}\ \bibnamefont {Matthewson}},\ }\bibfield  {title} {\bibinfo {title} {{Could We Be Fooled about Phantom Crossing?}},\ }\href@noop {} {\  (\bibinfo {year} {2025})},\ \Eprint {https://arxiv.org/abs/2506.15091} {arXiv:2506.15091 [astro-ph.CO]} \BibitemShut {NoStop}%
\bibitem [{\citenamefont {Mishra}(2026)}]{Mishra:2026tzn}%
  \BibitemOpen
  \bibfield  {author} {\bibinfo {author} {\bibfnamefont {S.~S.}\ \bibnamefont {Mishra}},\ }\bibfield  {title} {\bibinfo {title} {{Effective Phantom Dark Energy: What Cosmological Reconstruction Does and Does Not Imply}},\ }\href@noop {} {\  (\bibinfo {year} {2026})},\ \Eprint {https://arxiv.org/abs/2605.27301} {arXiv:2605.27301 [astro-ph.CO]} \BibitemShut {NoStop}%
\bibitem [{\citenamefont {Amin}\ \emph {et~al.}(2012{\natexlab{a}})\citenamefont {Amin}, \citenamefont {Zukin},\ and\ \citenamefont {Bertschinger}}]{Amin:2011hu}%
  \BibitemOpen
  \bibfield  {author} {\bibinfo {author} {\bibfnamefont {M.~A.}\ \bibnamefont {Amin}}, \bibinfo {author} {\bibfnamefont {P.}~\bibnamefont {Zukin}},\ and\ \bibinfo {author} {\bibfnamefont {E.}~\bibnamefont {Bertschinger}},\ }\bibfield  {title} {\bibinfo {title} {{Scale-Dependent Growth from a Transition in Dark Energy Dynamics}},\ }\href {https://doi.org/10.1103/PhysRevD.85.103510} {\bibfield  {journal} {\bibinfo  {journal} {Phys. Rev. D}\ }\textbf {\bibinfo {volume} {85}},\ \bibinfo {pages} {103510} (\bibinfo {year} {2012}{\natexlab{a}})},\ \Eprint {https://arxiv.org/abs/1108.1793} {arXiv:1108.1793 [astro-ph.CO]} \BibitemShut {NoStop}%
\bibitem [{\citenamefont {Johnson}\ and\ \citenamefont {Kamionkowski}(2008)}]{Johnson:2008se}%
  \BibitemOpen
  \bibfield  {author} {\bibinfo {author} {\bibfnamefont {M.~C.}\ \bibnamefont {Johnson}}\ and\ \bibinfo {author} {\bibfnamefont {M.}~\bibnamefont {Kamionkowski}},\ }\bibfield  {title} {\bibinfo {title} {{Dynamical and Gravitational Instability of Oscillating-Field Dark Energy and Dark Matter}},\ }\href {https://doi.org/10.1103/PhysRevD.78.063010} {\bibfield  {journal} {\bibinfo  {journal} {Phys. Rev. D}\ }\textbf {\bibinfo {volume} {78}},\ \bibinfo {pages} {063010} (\bibinfo {year} {2008})},\ \Eprint {https://arxiv.org/abs/0805.1748} {arXiv:0805.1748 [astro-ph]} \BibitemShut {NoStop}%
\bibitem [{\citenamefont {Amin}\ \emph {et~al.}(2012{\natexlab{b}})\citenamefont {Amin}, \citenamefont {Easther}, \citenamefont {Finkel}, \citenamefont {Flauger},\ and\ \citenamefont {Hertzberg}}]{Amin:2011hj}%
  \BibitemOpen
  \bibfield  {author} {\bibinfo {author} {\bibfnamefont {M.~A.}\ \bibnamefont {Amin}}, \bibinfo {author} {\bibfnamefont {R.}~\bibnamefont {Easther}}, \bibinfo {author} {\bibfnamefont {H.}~\bibnamefont {Finkel}}, \bibinfo {author} {\bibfnamefont {R.}~\bibnamefont {Flauger}},\ and\ \bibinfo {author} {\bibfnamefont {M.~P.}\ \bibnamefont {Hertzberg}},\ }\bibfield  {title} {\bibinfo {title} {{Oscillons After Inflation}},\ }\href {https://doi.org/10.1103/PhysRevLett.108.241302} {\bibfield  {journal} {\bibinfo  {journal} {Phys. Rev. Lett.}\ }\textbf {\bibinfo {volume} {108}},\ \bibinfo {pages} {241302} (\bibinfo {year} {2012}{\natexlab{b}})},\ \Eprint {https://arxiv.org/abs/1106.3335} {arXiv:1106.3335 [astro-ph.CO]} \BibitemShut {NoStop}%
\bibitem [{\citenamefont {Lozanov}\ and\ \citenamefont {Amin}(2018)}]{Lozanov:2017hjm}%
  \BibitemOpen
  \bibfield  {author} {\bibinfo {author} {\bibfnamefont {K.~D.}\ \bibnamefont {Lozanov}}\ and\ \bibinfo {author} {\bibfnamefont {M.~A.}\ \bibnamefont {Amin}},\ }\bibfield  {title} {\bibinfo {title} {{Self-resonance after inflation: oscillons, transients and radiation domination}},\ }\href {https://doi.org/10.1103/PhysRevD.97.023533} {\bibfield  {journal} {\bibinfo  {journal} {Phys. Rev. D}\ }\textbf {\bibinfo {volume} {97}},\ \bibinfo {pages} {023533} (\bibinfo {year} {2018})},\ \Eprint {https://arxiv.org/abs/1710.06851} {arXiv:1710.06851 [astro-ph.CO]} \BibitemShut {NoStop}%
\bibitem [{\citenamefont {Poulin}\ \emph {et~al.}(2019)\citenamefont {Poulin}, \citenamefont {Smith}, \citenamefont {Karwal},\ and\ \citenamefont {Kamionkowski}}]{Poulin:2018cxd}%
  \BibitemOpen
  \bibfield  {author} {\bibinfo {author} {\bibfnamefont {V.}~\bibnamefont {Poulin}}, \bibinfo {author} {\bibfnamefont {T.~L.}\ \bibnamefont {Smith}}, \bibinfo {author} {\bibfnamefont {T.}~\bibnamefont {Karwal}},\ and\ \bibinfo {author} {\bibfnamefont {M.}~\bibnamefont {Kamionkowski}},\ }\bibfield  {title} {\bibinfo {title} {{Early Dark Energy Can Resolve The Hubble Tension}},\ }\href {https://doi.org/10.1103/PhysRevLett.122.221301} {\bibfield  {journal} {\bibinfo  {journal} {Phys. Rev. Lett.}\ }\textbf {\bibinfo {volume} {122}},\ \bibinfo {pages} {221301} (\bibinfo {year} {2019})},\ \Eprint {https://arxiv.org/abs/1811.04083} {arXiv:1811.04083 [astro-ph.CO]} \BibitemShut {NoStop}%
\bibitem [{\citenamefont {Smith}\ \emph {et~al.}(2020)\citenamefont {Smith}, \citenamefont {Poulin},\ and\ \citenamefont {Amin}}]{Smith:2019ihp}%
  \BibitemOpen
  \bibfield  {author} {\bibinfo {author} {\bibfnamefont {T.~L.}\ \bibnamefont {Smith}}, \bibinfo {author} {\bibfnamefont {V.}~\bibnamefont {Poulin}},\ and\ \bibinfo {author} {\bibfnamefont {M.~A.}\ \bibnamefont {Amin}},\ }\bibfield  {title} {\bibinfo {title} {{Oscillating scalar fields and the Hubble tension: a resolution with novel signatures}},\ }\href {https://doi.org/10.1103/PhysRevD.101.063523} {\bibfield  {journal} {\bibinfo  {journal} {Phys. Rev. D}\ }\textbf {\bibinfo {volume} {101}},\ \bibinfo {pages} {063523} (\bibinfo {year} {2020})},\ \Eprint {https://arxiv.org/abs/1908.06995} {arXiv:1908.06995 [astro-ph.CO]} \BibitemShut {NoStop}%
\bibitem [{\citenamefont {Dodelson}\ \emph {et~al.}(2000)\citenamefont {Dodelson}, \citenamefont {Kaplinghat},\ and\ \citenamefont {Stewart}}]{Dodelson:2000jtt}%
  \BibitemOpen
  \bibfield  {author} {\bibinfo {author} {\bibfnamefont {S.}~\bibnamefont {Dodelson}}, \bibinfo {author} {\bibfnamefont {M.}~\bibnamefont {Kaplinghat}},\ and\ \bibinfo {author} {\bibfnamefont {E.}~\bibnamefont {Stewart}},\ }\bibfield  {title} {\bibinfo {title} {{Solving the Coincidence Problem : Tracking Oscillating Energy}},\ }\href {https://doi.org/10.1103/PhysRevLett.85.5276} {\bibfield  {journal} {\bibinfo  {journal} {Phys. Rev. Lett.}\ }\textbf {\bibinfo {volume} {85}},\ \bibinfo {pages} {5276} (\bibinfo {year} {2000})},\ \Eprint {https://arxiv.org/abs/astro-ph/0002360} {arXiv:astro-ph/0002360} \BibitemShut {NoStop}%
\bibitem [{\citenamefont {Kallosh}\ \emph {et~al.}(2013)\citenamefont {Kallosh}, \citenamefont {Linde},\ and\ \citenamefont {Roest}}]{Kallosh:2013yoa}%
  \BibitemOpen
  \bibfield  {author} {\bibinfo {author} {\bibfnamefont {R.}~\bibnamefont {Kallosh}}, \bibinfo {author} {\bibfnamefont {A.}~\bibnamefont {Linde}},\ and\ \bibinfo {author} {\bibfnamefont {D.}~\bibnamefont {Roest}},\ }\bibfield  {title} {\bibinfo {title} {{Superconformal Inflationary $\alpha$-Attractors}},\ }\href {https://doi.org/10.1007/JHEP11(2013)198} {\bibfield  {journal} {\bibinfo  {journal} {JHEP}\ }\textbf {\bibinfo {volume} {11}}\bibfield  {number} {\bibinfo  {number} { (11)},\ \bibinfo {pages} {198}},\ }\Eprint {https://arxiv.org/abs/1311.0472} {arXiv:1311.0472 [hep-th]} \BibitemShut {NoStop}%
\bibitem [{\citenamefont {Silverstein}\ and\ \citenamefont {Westphal}(2008)}]{Silverstein:2008sg}%
  \BibitemOpen
  \bibfield  {author} {\bibinfo {author} {\bibfnamefont {E.}~\bibnamefont {Silverstein}}\ and\ \bibinfo {author} {\bibfnamefont {A.}~\bibnamefont {Westphal}},\ }\bibfield  {title} {\bibinfo {title} {{Monodromy in the CMB: Gravity Waves and String Inflation}},\ }\href {https://doi.org/10.1103/PhysRevD.78.106003} {\bibfield  {journal} {\bibinfo  {journal} {Phys. Rev. D}\ }\textbf {\bibinfo {volume} {78}},\ \bibinfo {pages} {106003} (\bibinfo {year} {2008})},\ \Eprint {https://arxiv.org/abs/0803.3085} {arXiv:0803.3085 [hep-th]} \BibitemShut {NoStop}%
\bibitem [{\citenamefont {Andriot}(2026)}]{Andriot:2026lac}%
  \BibitemOpen
  \bibfield  {author} {\bibinfo {author} {\bibfnamefont {D.}~\bibnamefont {Andriot}},\ }\bibfield  {title} {\bibinfo {title} {{Dark energy from string theory: an introductory review}},\ }\href@noop {} {\bibfield  {journal} {\bibinfo  {journal} {arXiv e-prints}\ } (\bibinfo {year} {2026})},\ \Eprint {https://arxiv.org/abs/2603.25797} {arXiv:2603.25797 [hep-th]} \BibitemShut {NoStop}%
\bibitem [{Note1()}]{Note1}%
  \BibitemOpen
  \bibinfo {note} {In Appendix~B, we also present the results with a flat prior on the field parameters $(\log _{10}(\protect \mathcal {V}/H^2_0 m_{\protect \rm pl}^2), \log _{10} (M/m_{\protect \text {Pl}}))$.}\BibitemShut {Stop}%
\bibitem [{\citenamefont {Blas}\ \emph {et~al.}(2011)\citenamefont {Blas}, \citenamefont {Lesgourgues},\ and\ \citenamefont {Tram}}]{Blas:2011rf}%
  \BibitemOpen
  \bibfield  {author} {\bibinfo {author} {\bibfnamefont {D.}~\bibnamefont {Blas}}, \bibinfo {author} {\bibfnamefont {J.}~\bibnamefont {Lesgourgues}},\ and\ \bibinfo {author} {\bibfnamefont {T.}~\bibnamefont {Tram}},\ }\bibfield  {title} {\bibinfo {title} {{The Cosmic Linear Anisotropy Solving System (CLASS) II: Approximation schemes}},\ }\href {https://doi.org/10.1088/1475-7516/2011/07/034} {\bibfield  {journal} {\bibinfo  {journal} {JCAP}\ }\textbf {\bibinfo {volume} {07}},\ \bibinfo {pages} {034}},\ \Eprint {https://arxiv.org/abs/1104.2933} {arXiv:1104.2933 [astro-ph.CO]} \BibitemShut {NoStop}%
\bibitem [{\citenamefont {Rubin}\ \emph {et~al.}(2025)\citenamefont {Rubin} \emph {et~al.}}]{Rubin:2023jdq}%
  \BibitemOpen
  \bibfield  {author} {\bibinfo {author} {\bibfnamefont {D.}~\bibnamefont {Rubin}} \emph {et~al.},\ }\bibfield  {title} {\bibinfo {title} {{Union Through UNITY: Cosmology with 2,000 SNe Using a Unified Bayesian Framework}},\ }\href {https://doi.org/10.3847/1538-4357/adc0a5} {\bibfield  {journal} {\bibinfo  {journal} {Astrophys. J.}\ }\textbf {\bibinfo {volume} {986}},\ \bibinfo {pages} {231} (\bibinfo {year} {2025})},\ \Eprint {https://arxiv.org/abs/2311.12098} {arXiv:2311.12098 [astro-ph.CO]} \BibitemShut {NoStop}%
\bibitem [{\citenamefont {Rosenberg}\ \emph {et~al.}(2022)\citenamefont {Rosenberg}, \citenamefont {Gratton},\ and\ \citenamefont {Efstathiou}}]{Rosenberg:2022sdy}%
  \BibitemOpen
  \bibfield  {author} {\bibinfo {author} {\bibfnamefont {E.}~\bibnamefont {Rosenberg}}, \bibinfo {author} {\bibfnamefont {S.}~\bibnamefont {Gratton}},\ and\ \bibinfo {author} {\bibfnamefont {G.}~\bibnamefont {Efstathiou}},\ }\bibfield  {title} {\bibinfo {title} {{CMB power spectra and cosmological parameters from Planck PR4 with CamSpec}},\ }\href {https://doi.org/10.1093/mnras/stac2744} {\bibfield  {journal} {\bibinfo  {journal} {Mon. Not. Roy. Astron. Soc.}\ }\textbf {\bibinfo {volume} {517}},\ \bibinfo {pages} {4620} (\bibinfo {year} {2022})},\ \Eprint {https://arxiv.org/abs/2205.10869} {arXiv:2205.10869 [astro-ph.CO]} \BibitemShut {NoStop}%
\bibitem [{\citenamefont {Aghanim}\ \emph {et~al.}(2020{\natexlab{b}})\citenamefont {Aghanim} \emph {et~al.}}]{Planck:2019nip}%
  \BibitemOpen
  \bibfield  {author} {\bibinfo {author} {\bibfnamefont {N.}~\bibnamefont {Aghanim}} \emph {et~al.} (\bibinfo {collaboration} {Planck}),\ }\bibfield  {title} {\bibinfo {title} {{Planck 2018 results. V. CMB power spectra and likelihoods}},\ }\href {https://doi.org/10.1051/0004-6361/201936386} {\bibfield  {journal} {\bibinfo  {journal} {Astron. Astrophys.}\ }\textbf {\bibinfo {volume} {641}},\ \bibinfo {pages} {A5} (\bibinfo {year} {2020}{\natexlab{b}})},\ \Eprint {https://arxiv.org/abs/1907.12875} {arXiv:1907.12875 [astro-ph.CO]} \BibitemShut {NoStop}%
\bibitem [{\citenamefont {Lemos}\ and\ \citenamefont {Lewis}(2023)}]{Lemos:2023xhs}%
  \BibitemOpen
  \bibfield  {author} {\bibinfo {author} {\bibfnamefont {P.}~\bibnamefont {Lemos}}\ and\ \bibinfo {author} {\bibfnamefont {A.}~\bibnamefont {Lewis}},\ }\bibfield  {title} {\bibinfo {title} {{CMB constraints on the early Universe independent of late-time cosmology}},\ }\href {https://doi.org/10.1103/PhysRevD.107.103505} {\bibfield  {journal} {\bibinfo  {journal} {Phys. Rev. D}\ }\textbf {\bibinfo {volume} {107}},\ \bibinfo {pages} {103505} (\bibinfo {year} {2023})},\ \Eprint {https://arxiv.org/abs/2302.12911} {arXiv:2302.12911 [astro-ph.CO]} \BibitemShut {NoStop}%
\bibitem [{Note2()}]{Note2}%
  \BibitemOpen
  \bibinfo {note} {For the compressed CMB analysis, $A_s, n_s, \tau _\protect \text {reio}$ are not sampled because the compressed likelihood only constrains the geometric and physical density parameters.}\BibitemShut {Stop}%
\bibitem [{\citenamefont {Torrado}\ and\ \citenamefont {Lewis}(2021)}]{Torrado:2020dgo}%
  \BibitemOpen
  \bibfield  {author} {\bibinfo {author} {\bibfnamefont {J.}~\bibnamefont {Torrado}}\ and\ \bibinfo {author} {\bibfnamefont {A.}~\bibnamefont {Lewis}},\ }\bibfield  {title} {\bibinfo {title} {{Cobaya: Code for Bayesian Analysis of hierarchical physical models}},\ }\href {https://doi.org/10.1088/1475-7516/2021/05/057} {\bibfield  {journal} {\bibinfo  {journal} {JCAP}\ }\textbf {\bibinfo {volume} {05}},\ \bibinfo {pages} {057}},\ \Eprint {https://arxiv.org/abs/2005.05290} {arXiv:2005.05290 [astro-ph.IM]} \BibitemShut {NoStop}%
\bibitem [{\citenamefont {Cartis}\ \emph {et~al.}(2018)\citenamefont {Cartis}, \citenamefont {Fiala}, \citenamefont {Marteau},\ and\ \citenamefont {Roberts}}]{Cartis:2018xum}%
  \BibitemOpen
  \bibfield  {author} {\bibinfo {author} {\bibfnamefont {C.}~\bibnamefont {Cartis}}, \bibinfo {author} {\bibfnamefont {J.}~\bibnamefont {Fiala}}, \bibinfo {author} {\bibfnamefont {B.}~\bibnamefont {Marteau}},\ and\ \bibinfo {author} {\bibfnamefont {L.}~\bibnamefont {Roberts}},\ }\bibfield  {title} {\bibinfo {title} {{Improving the Flexibility and Robustness of Model-Based Derivative-Free Optimization Solvers}},\ }\href@noop {} {\  (\bibinfo {year} {2018})},\ \Eprint {https://arxiv.org/abs/1804.00154} {arXiv:1804.00154 [math.OC]} \BibitemShut {NoStop}%
\bibitem [{\citenamefont {Cartis}\ \emph {et~al.}(2021)\citenamefont {Cartis}, \citenamefont {Roberts},\ and\ \citenamefont {Sheridan-Methven}}]{Cartis:2018jxl}%
  \BibitemOpen
  \bibfield  {author} {\bibinfo {author} {\bibfnamefont {C.}~\bibnamefont {Cartis}}, \bibinfo {author} {\bibfnamefont {L.}~\bibnamefont {Roberts}},\ and\ \bibinfo {author} {\bibfnamefont {O.}~\bibnamefont {Sheridan-Methven}},\ }\bibfield  {title} {\bibinfo {title} {{Escaping local minima with local derivative-free methods: a numerical investigation}},\ }\href {https://doi.org/10.1080/02331934.2021.1883015} {\bibfield  {journal} {\bibinfo  {journal} {Optimization}\ }\textbf {\bibinfo {volume} {71}},\ \bibinfo {pages} {2343} (\bibinfo {year} {2021})},\ \Eprint {https://arxiv.org/abs/1812.11343} {arXiv:1812.11343 [math.OC]} \BibitemShut {NoStop}%
\bibitem [{\citenamefont {Holm}\ \emph {et~al.}(2024)\citenamefont {Holm}, \citenamefont {Nygaard}, \citenamefont {Dakin}, \citenamefont {Hannestad},\ and\ \citenamefont {Tram}}]{Holm:2023uwa}%
  \BibitemOpen
  \bibfield  {author} {\bibinfo {author} {\bibfnamefont {E.~B.}\ \bibnamefont {Holm}}, \bibinfo {author} {\bibfnamefont {A.}~\bibnamefont {Nygaard}}, \bibinfo {author} {\bibfnamefont {J.}~\bibnamefont {Dakin}}, \bibinfo {author} {\bibfnamefont {S.}~\bibnamefont {Hannestad}},\ and\ \bibinfo {author} {\bibfnamefont {T.}~\bibnamefont {Tram}},\ }\bibfield  {title} {\bibinfo {title} {{PROSPECT: a profile likelihood code for frequentist cosmological parameter inference}},\ }\href {https://doi.org/10.1093/mnras/stae2555} {\bibfield  {journal} {\bibinfo  {journal} {Mon. Not. Roy. Astron. Soc.}\ }\textbf {\bibinfo {volume} {535}},\ \bibinfo {pages} {3686} (\bibinfo {year} {2024})},\ \Eprint {https://arxiv.org/abs/2312.02972} {arXiv:2312.02972 [astro-ph.CO]} \BibitemShut {NoStop}%
\bibitem [{Note3()}]{Note3}%
  \BibitemOpen
  \bibinfo {note} {The best-fit values are within an order of magnitude of those considered in a related 2011 work by one of us~\cite {Amin:2011hu}. In that case, the motivation was a ``what if'' exploration of the observational consequences of such behavior, rather than a response to DESI data. Separately, it is interesting that Ref.~\cite {Shiu:2026edl}, combining the requirement of evolving dark energy with the axionic Weak Gravity Conjecture~\cite {Heidenreich:2015wga}, obtains a similar scale, $m\sim 10^2H_0$.}\BibitemShut {Stop}%
\bibitem [{Note4()}]{Note4}%
  \BibitemOpen
  \bibinfo {note} {Due to the non-Gaussian and multimodal posteriors, we use the definition DIC$=\protect \bar {\chi ^2}+p_V$ with $p_V = {\protect \rm Var}[\chi ^2]/2$~\cite {Gelman2004BDA}.}\BibitemShut {Stop}%
\bibitem [{\citenamefont {Chevallier}\ and\ \citenamefont {Polarski}(2001)}]{Chevallier:2000qy}%
  \BibitemOpen
  \bibfield  {author} {\bibinfo {author} {\bibfnamefont {M.}~\bibnamefont {Chevallier}}\ and\ \bibinfo {author} {\bibfnamefont {D.}~\bibnamefont {Polarski}},\ }\bibfield  {title} {\bibinfo {title} {{Accelerating universes with scaling dark matter}},\ }\href {https://doi.org/10.1142/S0218271801000822} {\bibfield  {journal} {\bibinfo  {journal} {Int. J. Mod. Phys. D}\ }\textbf {\bibinfo {volume} {10}},\ \bibinfo {pages} {213} (\bibinfo {year} {2001})},\ \Eprint {https://arxiv.org/abs/gr-qc/0009008} {arXiv:gr-qc/0009008} \BibitemShut {NoStop}%
\bibitem [{\citenamefont {Linder}(2003)}]{Linder:2002et}%
  \BibitemOpen
  \bibfield  {author} {\bibinfo {author} {\bibfnamefont {E.~V.}\ \bibnamefont {Linder}},\ }\bibfield  {title} {\bibinfo {title} {{Exploring the expansion history of the universe}},\ }\href {https://doi.org/10.1103/PhysRevLett.90.091301} {\bibfield  {journal} {\bibinfo  {journal} {Phys. Rev. Lett.}\ }\textbf {\bibinfo {volume} {90}},\ \bibinfo {pages} {091301} (\bibinfo {year} {2003})},\ \Eprint {https://arxiv.org/abs/astro-ph/0208512} {arXiv:astro-ph/0208512} \BibitemShut {NoStop}%
\bibitem [{Note5()}]{Note5}%
  \BibitemOpen
  \bibinfo {note} {When the CMB information is replaced by a BBN prior, we find comparable preferences for our model and for the $w_0w_a$ parameterization.}\BibitemShut {Stop}%
\bibitem [{Note6()}]{Note6}%
  \BibitemOpen
  \bibinfo {note} {Our quintessence field is effectively massless during inflation, so $ \phi _{\protect \rm ini}\sim H_{\protect \rm inf}\ll M$ if it started at $\phi =0$ before $\sim 60$ e-folds of inflation began. However, since $\protect \mathcal {V}\sim m_{\protect \rm pl}^2H_0^2\ll H_{\protect \rm inf}^4\ll E_{\protect \rm inf}^4$, one could argue the field could have started anywhere.}\BibitemShut {Stop}%
\bibitem [{\citenamefont {Emami}\ \emph {et~al.}(2016)\citenamefont {Emami}, \citenamefont {Grin}, \citenamefont {Pradler}, \citenamefont {Raccanelli},\ and\ \citenamefont {Kamionkowski}}]{Emami:2016mrt}%
  \BibitemOpen
  \bibfield  {author} {\bibinfo {author} {\bibfnamefont {R.}~\bibnamefont {Emami}}, \bibinfo {author} {\bibfnamefont {D.}~\bibnamefont {Grin}}, \bibinfo {author} {\bibfnamefont {J.}~\bibnamefont {Pradler}}, \bibinfo {author} {\bibfnamefont {A.}~\bibnamefont {Raccanelli}},\ and\ \bibinfo {author} {\bibfnamefont {M.}~\bibnamefont {Kamionkowski}},\ }\bibfield  {title} {\bibinfo {title} {{Cosmological tests of an axiverse-inspired quintessence field}},\ }\href {https://doi.org/10.1103/PhysRevD.93.123005} {\bibfield  {journal} {\bibinfo  {journal} {Phys. Rev. D}\ }\textbf {\bibinfo {volume} {93}},\ \bibinfo {pages} {123005} (\bibinfo {year} {2016})},\ \Eprint {https://arxiv.org/abs/1603.04851} {arXiv:1603.04851 [astro-ph.CO]} \BibitemShut {NoStop}%
\bibitem [{\citenamefont {Shiu}\ \emph {et~al.}(2026)\citenamefont {Shiu}, \citenamefont {Tonioni},\ and\ \citenamefont {Tran}}]{Shiu:2026edl}%
  \BibitemOpen
  \bibfield  {author} {\bibinfo {author} {\bibfnamefont {G.}~\bibnamefont {Shiu}}, \bibinfo {author} {\bibfnamefont {F.}~\bibnamefont {Tonioni}},\ and\ \bibinfo {author} {\bibfnamefont {H.~V.}\ \bibnamefont {Tran}},\ }\bibfield  {title} {\bibinfo {title} {{Bounding axion dark energy}},\ }\href@noop {} {\  (\bibinfo {year} {2026})},\ \Eprint {https://arxiv.org/abs/2604.09141} {arXiv:2604.09141 [astro-ph.CO]} \BibitemShut {NoStop}%
\bibitem [{\citenamefont {Smith}\ \emph {et~al.}(2023)\citenamefont {Smith}, \citenamefont {Giblin}, \citenamefont {Amin}, \citenamefont {Gerhardinger}, \citenamefont {Florio}, \citenamefont {Cerep},\ and\ \citenamefont {Daniels}}]{Smith:2023fob}%
  \BibitemOpen
  \bibfield  {author} {\bibinfo {author} {\bibfnamefont {T.~L.}\ \bibnamefont {Smith}}, \bibinfo {author} {\bibfnamefont {J.~T.}\ \bibnamefont {Giblin}, \bibfnamefont {Jr.}}, \bibinfo {author} {\bibfnamefont {M.~A.}\ \bibnamefont {Amin}}, \bibinfo {author} {\bibfnamefont {M.}~\bibnamefont {Gerhardinger}}, \bibinfo {author} {\bibfnamefont {E.}~\bibnamefont {Florio}}, \bibinfo {author} {\bibfnamefont {M.}~\bibnamefont {Cerep}},\ and\ \bibinfo {author} {\bibfnamefont {S.}~\bibnamefont {Daniels}},\ }\bibfield  {title} {\bibinfo {title} {{Novel integrated Sachs-Wolfe effect from early dark energy}},\ }\href {https://doi.org/10.1103/PhysRevD.108.123534} {\bibfield  {journal} {\bibinfo  {journal} {Phys. Rev. D}\ }\textbf {\bibinfo {volume} {108}},\ \bibinfo {pages} {123534} (\bibinfo {year} {2023})},\ \Eprint {https://arxiv.org/abs/2304.02028} {arXiv:2304.02028 [astro-ph.CO]} \BibitemShut {NoStop}%
\bibitem [{\citenamefont {Heidenreich}\ \emph {et~al.}(2015)\citenamefont {Heidenreich}, \citenamefont {Reece},\ and\ \citenamefont {Rudelius}}]{Heidenreich:2015wga}%
  \BibitemOpen
  \bibfield  {author} {\bibinfo {author} {\bibfnamefont {B.}~\bibnamefont {Heidenreich}}, \bibinfo {author} {\bibfnamefont {M.}~\bibnamefont {Reece}},\ and\ \bibinfo {author} {\bibfnamefont {T.}~\bibnamefont {Rudelius}},\ }\bibfield  {title} {\bibinfo {title} {{Weak Gravity Strongly Constrains Large-Field Axion Inflation}},\ }\href {https://doi.org/10.1007/JHEP12(2015)108} {\bibfield  {journal} {\bibinfo  {journal} {JHEP}\ }\textbf {\bibinfo {volume} {12}},\ \bibinfo {pages} {108}},\ \Eprint {https://arxiv.org/abs/1506.03447} {arXiv:1506.03447 [hep-th]} \BibitemShut {NoStop}%
\bibitem [{\citenamefont {Gelman}\ \emph {et~al.}(2004)\citenamefont {Gelman}, \citenamefont {Carlin}, \citenamefont {Stern},\ and\ \citenamefont {Rubin}}]{Gelman2004BDA}%
  \BibitemOpen
  \bibfield  {author} {\bibinfo {author} {\bibfnamefont {A.}~\bibnamefont {Gelman}}, \bibinfo {author} {\bibfnamefont {J.~B.}\ \bibnamefont {Carlin}}, \bibinfo {author} {\bibfnamefont {H.~S.}\ \bibnamefont {Stern}},\ and\ \bibinfo {author} {\bibfnamefont {D.~B.}\ \bibnamefont {Rubin}},\ }\href@noop {} {\emph {\bibinfo {title} {Bayesian Data Analysis}}},\ \bibinfo {edition} {2nd}\ ed.\ (\bibinfo  {publisher} {Chapman and Hall/CRC},\ \bibinfo {year} {2004})\BibitemShut {NoStop}%
\bibitem [{\citenamefont {Kallosh}\ and\ \citenamefont {Linde}(2013)}]{Kallosh:2013hoa}%
  \BibitemOpen
  \bibfield  {author} {\bibinfo {author} {\bibfnamefont {R.}~\bibnamefont {Kallosh}}\ and\ \bibinfo {author} {\bibfnamefont {A.}~\bibnamefont {Linde}},\ }\bibfield  {title} {\bibinfo {title} {{Universality Class in Conformal Inflation}},\ }\href {https://doi.org/10.1088/1475-7516/2013/07/002} {\bibfield  {journal} {\bibinfo  {journal} {JCAP}\ }\textbf {\bibinfo {volume} {07}},\ \bibinfo {pages} {002}},\ \Eprint {https://arxiv.org/abs/1306.5220} {arXiv:1306.5220 [hep-th]} \BibitemShut {NoStop}%
\bibitem [{\citenamefont {Scolnic}\ \emph {et~al.}(2022)\citenamefont {Scolnic} \emph {et~al.}}]{Scolnic:2021amr}%
  \BibitemOpen
  \bibfield  {author} {\bibinfo {author} {\bibfnamefont {D.}~\bibnamefont {Scolnic}} \emph {et~al.},\ }\bibfield  {title} {\bibinfo {title} {{The Pantheon+ Analysis: The Full Data Set and Light-curve Release}},\ }\href {https://doi.org/10.3847/1538-4357/ac8b7a} {\bibfield  {journal} {\bibinfo  {journal} {Astrophys. J.}\ }\textbf {\bibinfo {volume} {938}},\ \bibinfo {pages} {113} (\bibinfo {year} {2022})},\ \Eprint {https://arxiv.org/abs/2112.03863} {arXiv:2112.03863 [astro-ph.CO]} \BibitemShut {NoStop}%
\bibitem [{\citenamefont {Brout}\ \emph {et~al.}(2022)\citenamefont {Brout} \emph {et~al.}}]{Brout:2022vxf}%
  \BibitemOpen
  \bibfield  {author} {\bibinfo {author} {\bibfnamefont {D.}~\bibnamefont {Brout}} \emph {et~al.},\ }\bibfield  {title} {\bibinfo {title} {{The Pantheon+ Analysis: Cosmological Constraints}},\ }\href {https://doi.org/10.3847/1538-4357/ac8e04} {\bibfield  {journal} {\bibinfo  {journal} {Astrophys. J.}\ }\textbf {\bibinfo {volume} {938}},\ \bibinfo {pages} {110} (\bibinfo {year} {2022})},\ \Eprint {https://arxiv.org/abs/2202.04077} {arXiv:2202.04077 [astro-ph.CO]} \BibitemShut {NoStop}%
\bibitem [{\citenamefont {Popovic}\ \emph {et~al.}(2025)\citenamefont {Popovic} \emph {et~al.}}]{DES:2025sig}%
  \BibitemOpen
  \bibfield  {author} {\bibinfo {author} {\bibfnamefont {B.}~\bibnamefont {Popovic}} \emph {et~al.} (\bibinfo {collaboration} {DES}),\ }\bibfield  {title} {\bibinfo {title} {{The Dark Energy Survey Supernova Program: A Reanalysis Of Cosmology Results And Evidence For Evolving Dark Energy With An Updated Type Ia Supernova Calibration}}\ }\href {https://doi.org/10.1093/mnras/stag632} {10.1093/mnras/stag632} (\bibinfo {year} {2025}),\ \Eprint {https://arxiv.org/abs/2511.07517} {arXiv:2511.07517 [astro-ph.CO]} \BibitemShut {NoStop}%
\end{thebibliography}%
\appendix
\section{Appendices}

\section{A. Full posterior distribution}
\begin{figure*}[t]
  \centering
  \includegraphics[width=\textwidth]{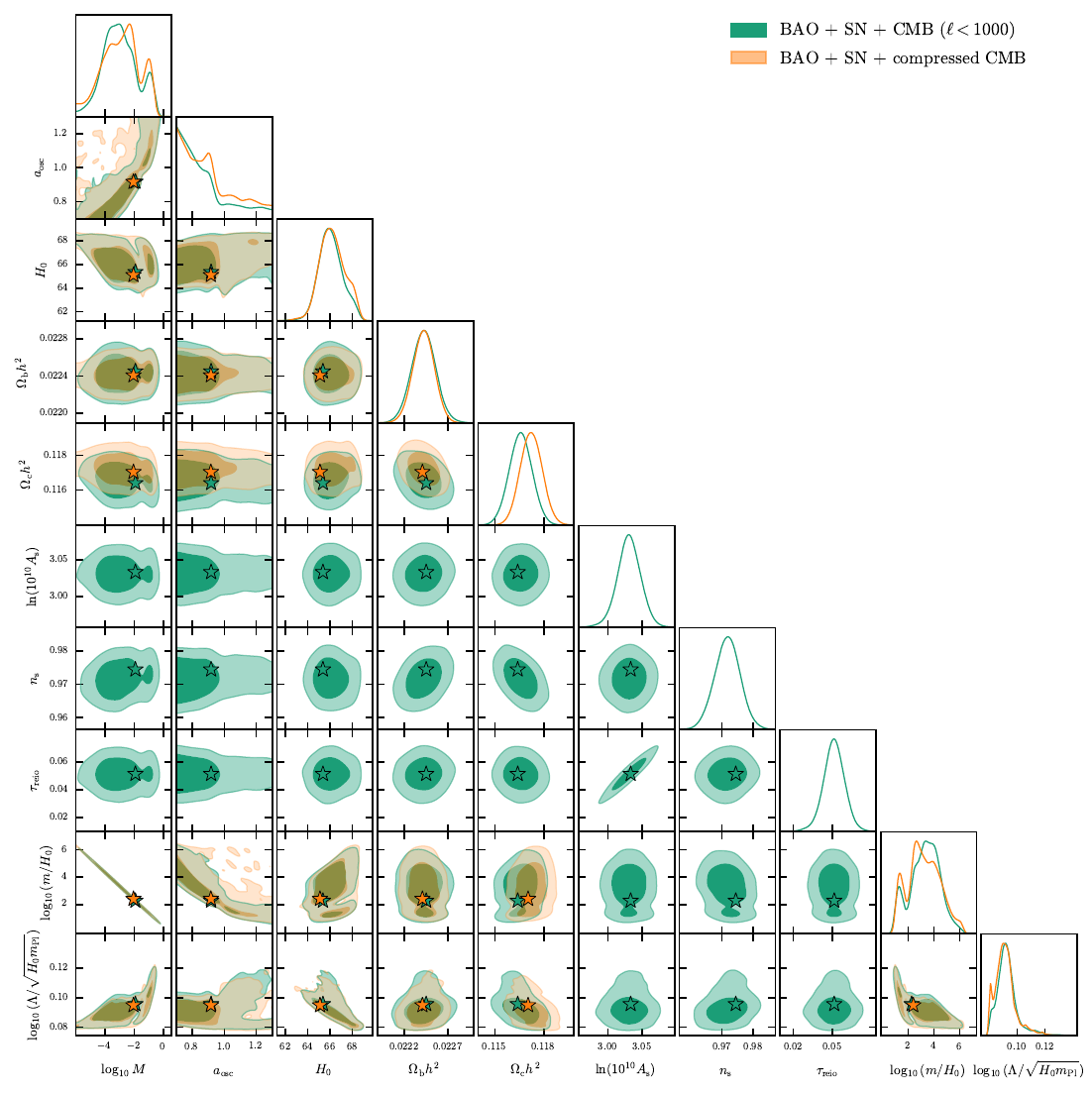}
  \caption{Marginalized one- and two-dimensional posterior distributions for the late-time oscillating quintessence model used in the main text. We show results for DESI DR2 BAO + Union3 supernovae combined with CMB spectra restricted to $\ell<1000$ and with the compressed CMB likelihood. Filled contours denote the 68\% and 95\% credible regions, and stars mark the best-fit points.}
  \label{fig:full_posterior_triangle}
\end{figure*}

This appendix summarizes the priors and posterior constraints used in the main analysis.
The priors adopted in our analysis are summarized in Table~\ref{tab:priors}.
The full marginalized constraints are given in Table~\ref{tab:full_posterior_constraints}, and the corresponding one- and two-dimensional posterior distributions are shown in Fig.~\ref{fig:full_posterior_triangle}.

\begin{table}[t]
  \centering
  \begin{tabular}{|c|c|} \hline
    Parameter & Prior \\ \hline
    $\log_{10} (M/m_{\rm Pl})$ & $[-6,-1]$ \\
    $a_{\mathrm{osc}}$ & $[0.7,1.3]$ \\ \hline
    $\ln(10^{10} A_{\mathrm{s}})$ & $[1.61,3.91]$ \\
    $n_{\mathrm{s}}$ & $[0.8,1.2]$ \\
    $H_0$ & $[20,100]$ \\
    $\Omega_{\mathrm{b}} h^2$ & $[0.005,0.1]$ \\
    $\Omega_{\mathrm{c}} h^2$ & $[0.001,0.99]$ \\
    $\tau_{\mathrm{reio}}$ & $[0.01,0.8]$ \\ \hline
  \end{tabular}
  \caption{Flat priors adopted for the sampled parameters in our analysis.}
  \label{tab:priors}
\end{table}

\begin{table*}
    \centering
\begin{tabular}{|c|cc|}
\hline
Parameter & BAO + SN + CMB ($\ell<1000$) & BAO + SN + compressed CMB \\
\hline
$\log_{10}(M/m_{\rm Pl})$ & $-2.9^{+2.0}_{-1.4}\;(-1.928)$ & $-2.9^{+2.0}_{-1.3}\;(-2.06)$ \\
$a_\mathrm{osc}$ & $< 1.21\;(0.9174)$ & $< 1.22\;(0.9161)$ \\
$H_0$ & $66.09^{+0.94}_{-1.2}\;(65.36)$ & $66.22^{+0.98}_{-1.3}\;(65.1)$ \\
$\Omega_\mathrm{b} h^2$ & $0.02243\pm 0.00014\;(0.02245)$ & $0.02242\pm 0.00012\;(0.02241)$ \\
$\Omega_\mathrm{c} h^2$ & $0.11658\pm 0.00069\;(0.1164)$ & $0.11723\pm 0.00067\;(0.117)$ \\
$\ln(10^{10} A_\mathrm{s})$ & $3.031\pm 0.017\;(3.033)$ & -- \\
$n_\mathrm{s}$ & $0.9717\pm 0.0040\;(0.9744)$ & -- \\
$\tau_\mathrm{reio}$ & $0.0511\pm 0.0081\;(0.05132)$ & -- \\ \hline
$\log_{10}(m/H_0)$ & $3.2^{+1.3}_{-2.0}\;(2.27)$ & $3.2^{+1.3}_{-2.0}\;(2.4)$ \\
$\log_{10}(\Lambda/\sqrt{H_0 m_\mathrm{Pl}})$ & $0.0932^{+0.0046}_{-0.0078}\;(0.09567)$ & $0.0917^{+0.0047}_{-0.011}\;(0.0949)$ \\
\hline
$\chi^2-\chi^2_{\Lambda{\rm CDM}}$ & -9.327 & -9.189 \\
${\rm DIC} - {\rm DIC}_{\Lambda{\rm CDM}}$ & -4.492 & -3.312 \\
\hline
\end{tabular}
    \caption{Marginalized constraints for the late-time oscillating quintessence model used in the main text. Quoted uncertainties denote 68\% credible intervals, and values in parentheses give the corresponding best-fit points. The two columns correspond to DESI DR2 BAO + Union3 supernovae combined either with the Planck PR4 CMB spectra restricted to $\ell<1000$ or with the compressed PR4 CMB likelihood.}
    \label{tab:full_posterior_constraints}
\end{table*}

The two CMB treatments lead to broadly consistent posteriors for both the cosmological parameters and the scalar-field parameters.
In both cases the marginalized posterior for $a_{\rm osc}$ peaks at the region with oscillation in the past $a_{\rm osc} < 1$.

\section{B. Alternative parameterization}
In the main analysis we use a flat prior on $a_{\rm osc}$, which directly targets the epoch at which the field first crosses the minimum and begins oscillating.
As a robustness check, we also consider a parameterization in terms of the underlying field parameters.
Specifically, we impose flat priors on $\log_{10}(M/m_{\rm Pl})\in[-6,3]$ and $\log_{10}(\Lambda/\sqrt{H_0m_{\rm Pl}})\in[0,0.3]$, where $\Lambda^4 \equiv \mathcal{V}$.
This range covers solutions whose oscillations begin earlier than the lower bound $a_{\rm osc}=0.7$ used in the main analysis, as well as thawing solutions whose first zero crossing would occur far in the future.
Although we modified \texttt{CLASS} to estimate $a_{\rm osc}$ for future oscillation solutions, many points in this branch do not cross zero within the numerical integration range, implying $a_{\rm osc}\gtrsim 100$ for these samples.
In the corresponding chains, samples with $a_{\rm osc}>1$ account for $21.68\%$ of the posterior sample weight, so the posterior still predominantly favors models in which the onset of oscillations has already occurred.

\begin{figure}[htp]
  \centering
  \includegraphics[width=\linewidth]{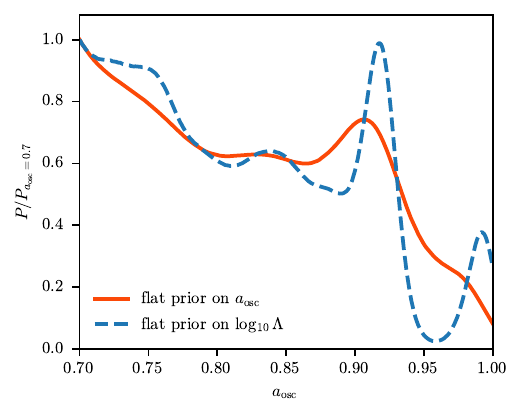}
  \caption{Marginalized posterior for $a_{\rm osc}$ restricted to $a_{\rm osc}<1$, comparing the main analysis with a flat prior on $a_{\rm osc}$ to the alternative field parameter prior. The distributions are normalized to their value at $a_{\rm osc}=0.7$. The similar shapes show that the preference for recent oscillations in the past is not driven by the choice of sampling parameter.}
  \label{fig:a_osc_prior_comparison}
\end{figure}

The one-dimensional marginalized posterior for the $a_{\rm osc}<1$ branch is shown in \autoref{fig:a_osc_prior_comparison}.
The alternative parameterization gives a posterior shape similar to that obtained with the flat prior on $a_{\rm osc}$.
This comparison indicates that the preference for late-time oscillations is not an artifact of using $a_{\rm osc}$ as a sampling parameter.
We also emphasize that the likelihood profile in \autoref{fig:a_osc} is a frequentist result obtained by optimizing the remaining parameters at fixed $a_{\rm osc}$, and is therefore independent of the Bayesian prior used to sample the posterior.

\section{C. Alternative scalar-field potentials}
In this appendix we consider alternative scalar-field potentials that can produce similar late-time oscillatory behavior.
As a representative example, we consider the T-model potential~\cite{Kallosh:2013hoa},
\begin{equation}
    V(\phi)=\Lambda^4 \tanh^{2n}\left(\frac{|\phi|}{M}\right).
\end{equation}
This potential has the same qualitative structure as the model used in the main text: it approaches a plateau at large field values,
\begin{equation}
    V(|\phi|\gg M)\simeq \Lambda^4,
\end{equation}
while near the minimum it reduces to a power-law form,
\begin{equation}
    V(|\phi|\ll M)\simeq \Lambda^4 \left|\frac{\phi}{M}\right|^{2n}.
\end{equation}
Oscillatory dynamics in this class of potentials have also been discussed in the context of post-inflationary evolution~\cite{Lozanov:2017hjm}.

We study the cases $n=1$ and $n=2$, corresponding respectively to quadratic and quartic behavior near the minimum.
The field evolution in these models is numerically more demanding than in the fiducial potential, because the late-time oscillations are more rapid.
We therefore run the MCMC chains until they reach $R-1<0.05$ and search the best-fit points based on the MCMC results.
For $n=1$, the best-fit point has $a_{\rm osc}=0.915$, while for $n=2$ it has $a_{\rm osc}=0.911$.
Both values are below unity, indicating that the field begins oscillating in the past.
Relative to the best-fit $\Lambda$CDM model for the same data set, the corresponding improvements are $\Delta\chi^2=-8.2$ and $\Delta\chi^2=-7.7$, respectively.
These results indicate that the preference for recent oscillatory dynamics is not a special feature of the particular potential adopted in the main analysis.

\section{D. Results with alternative SNe Ia dataset}
As a further robustness check, we compare the marginalized posterior distribution of $a_{\rm osc}$ after replacing Union3 with alternative Type Ia supernova data sets, using either Pantheon+~\cite{Scolnic:2021amr,Brout:2022vxf} or the new DES-Dovekie likelihood~\cite{DES:2025sig}.
In each case we keep the DESI BAO data and compressed CMB prior fixed, so that the comparison isolates the impact of the supernova sample.

\begin{figure}[htp!]
  \centering
  \includegraphics[width=\linewidth]{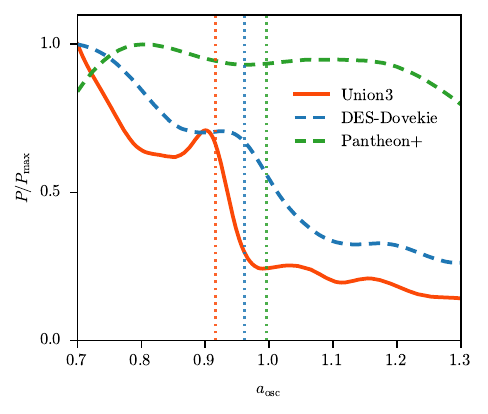}
  \caption{Marginalized posterior distributions for $a_{\rm osc}$ using DESI BAO + compressed CMB combined with different Type Ia supernova data sets, normalized to their respective maxima. The solid, dashed, and dash-dotted curves correspond to Union3, DES-Dovekie, and Pantheon+, respectively, and the vertical lines mark the corresponding best-fit values. The DES-Dovekie and Pantheon+ samples shift the preferred onset of oscillations closer to the present, but their best-fit values remain below unity.}
  \label{fig:aosc_alter_sn}
\end{figure}

The comparison is shown in \autoref{fig:aosc_alter_sn}.
For DES-Dovekie, the best-fit point is $a_{\rm osc}=0.961$, while for Pantheon+ it is $a_{\rm osc}=0.996$.
Both values are below unity, although to different degrees, indicating that the field still prefers to have crossed the minimum before today.
The preference is weakest for Pantheon+, where the best-fit onset is very close to the present epoch.
This is consistent with the broader background-level behavior of these supernova samples: even without invoking oscillations, Pantheon+ is relatively less favorable to dynamical dark energy than Union3 or DES-Dovekie.

\end{document}